\documentstyle[proceedings,psfig]{crckapb}
\def\kms{km~s$^{-1}$}
\def\Msun{M$_\odot$}
\begin{opening}
\title{Barium and Tc-poor S stars: Binary masqueraders among 
carbon stars}
\author{A. Jorissen}
\author{S. Van Eck}
\institute{Institut d'Astronomie et d'Astrophysique,\\
Universit\'e Libre de Bruxelles\\
Campus Plaine C.P.226, Boulevard du Triomphe,\\
B-1050 Bruxelles, 
Belgium}
\end{opening}
\runningtitle{Ba and Tc-poor S stars: Binary masqueraders} 
\begin{document}
\begin{abstract}
Our current understanding of the origin of barium and S stars is
briefly reviewed, based on new orbital elements and binary frequencies. 
\end{abstract}
\section{The relation of barium and S stars to carbon stars}
Since the last conference (IAU Coll. 106, {\it Evolution of Peculiar
Red Giants}, Johnson \& Zuckermann eds., 1989) 
devoted to chemically-peculiar red giants (PRGs), 
much progress has been made in understanding how barium 
and S stars relate to the other PRGs. The discovery of the binary 
nature of barium stars (McClure et al. 1980; McClure 1983) 
suggested from the beginning that mass transfer was likely to play a 
key role in the formation of the barium syndrome. 
As far as S stars are concerned, it has become clear that 
Tc-rich and Tc-poor S stars form two separate families with similar  
chemical peculiarities albeit of very different  origins (Iben \& Renzini 
1983; Little-Marenin et al. 1987; Jorissen \& Mayor 1988; Smith \&
Lambert 1988; Brown et al. 1990; Johnson 1992; Jorissen \& Mayor 1992;
Groenewegen 1993; Johnson et al. 1993; Jorissen et al. 1993; Ake, this
conference).  
Tc-rich (or `intrinsic') S stars are 
genuine thermally-pulsing AGB stars where the s-process operates in 
relation with the thermal pulses, and where the third dredge-up 
brings the freshly synthesized s-elements (including Tc) to the 
surface (e.g. Iben \& Renzini 1983; Sackmann \& Boothroyd 1991). 
By contrast, Tc-poor  (or 
`extrinsic') S stars are believed to be the cool descendants of barium 
stars.

\begin{figure}
\begin{center}
\leavevmode
\centerline{\psfig{file=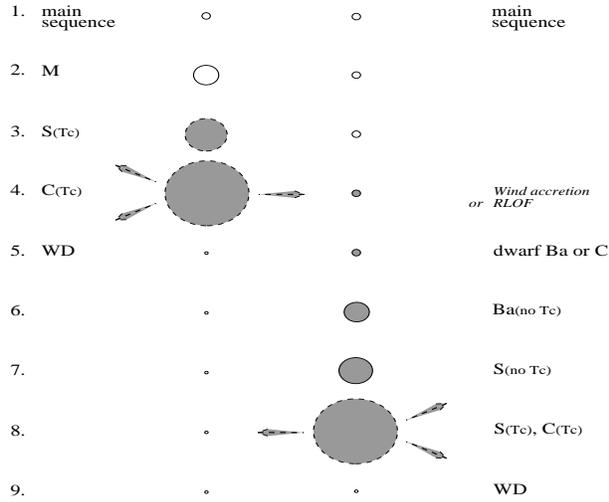,width=8cm,height=9cm}}
\end{center}
\caption[t]{
Relationship between several families of PRG stars. Grey symbols 
represent heavy-element-rich stars, and dashed boundaries indicate 
Tc-rich stars. The left column depicts the normal 
(i.e. not requiring binarity) M--S--C evolution on the AGB, 
whereas the right column 
represents the evolution of a companion star. Note in 
particular the possibility  that this companion itself evolves into a 
Tc-rich S star on the AGB, after having first shown up as an 
extrinsic S  star
} 
\end{figure}

Figure 1 summarizes our current understanding of the relationship 
between the different families of PRG stars. 
This general picture raises several questions, that will briefly be 
addressed in this paper:\\
\begin{enumerate}
\item Is binarity a necessary condition to produce a barium star? 
\item What is the mass transfer mode (wind accretion or RLOF?) 
responsible for their formation?
\item Do barium stars form as dwarfs or as giants?
\item Do barium stars evolve into Tc-poor S stars?
\item What is the relative frequency of Tc-rich and Tc-poor S stars?
\item Are the abundances in the mass-loser star
(i.e the AGB progenitor of the present white dwarf
companion) compatible with those presently observed 
in the barium or extrinsic S star?  
\end{enumerate}

We refer to Jorissen \& Boffin (1992), Han et al. (1995) and Busso et 
al. (1995) for a detailed discussion of item 6.
 
\section{Is binarity a necessary condition to form a barium star?}

To answer that question, {\it all}\ 27 barium stars with {\it strong} 
anomalies (i.e. Ba3, Ba4 or Ba5 on the scale devised by Warner 1965) 
south of $\delta = -25^{\circ}$ from the list of L\"u et al. (1983) 
have been monitored with the CORAVEL spectrovelocimeter (Baranne 
et al. 1979) since 1984. HD~19014 is the 
only star  in that sample that does not show any sign of binary 
motion. No detailed abundance analysis to confirm the barium nature 
of that star is available, unfortunately. For a fictitious population
of binaries
observed with the same time sampling and the same internal errors
as the real sample of barium stars, 
and having eccentricity and mass-function distributions
matching the observed ones,
a Monte-Carlo simulation yields a binary detection rate comprised between 96\%
(25.9/27) and 98\% (26.5/27), depending on whether the observed period
distribution is extrapolated or not towards periods as
long as $2\times 10^4$~d [see Jorissen et al. 1997 for more details].
{\it Binarity is thus a necessary 
condition to produce strong barium stars}.

In a comparison sample of 28 {\it mild} barium stars (i.e. with Ba1 
and Ba2 indices) randomly selected from the list of L\"u et al. 
(1983) and monitored in a similar way as the strong barium stars, 
23 (82\%) are definitely spectroscopic binaries, 2 (7\%) are 
probably binaries, and 3 (11\%; HD 50843, HD 95345, HD 119185) 
show no sign of radial velocity variations at the level 0.3 \kms\ 
r.m.s. after more than 10 y of monitoring.
Detailed spectroscopic abundance analyses
performed on HD 95345 (Sneden et al. 1981) and HD 119185 
(Za\v{c}s et al. 1996) confirm the existence of mild heavy-element 
overabundances ([s/Fe] = 0.2 to 0.3 dex) for these stars with constant
radial velocity.  
This frequency of constant stars is again consistent with the
binary detection rate predicted for that sample by a Monte-Carlo
simulation, provided
that the period distribution of mild barium stars extends up to
$2\times 10^4$~d.
In these conditions, there is no need to invoke any formation
mechanism other than mass transfer in a binary system 
to produce mild barium stars. On the contrary, 
an alternative formation scenario (like galactic fluctuations of the
s/Fe ratio; Williams 1975, Sneden et al. 1981, Edvardsson et al. 1993)
may be required to account for a population of 
non-binary stars found among {\it dwarf}
mild barium stars (North et al., this conference).  
 

\begin{figure}
\begin{center}
\leavevmode
\centerline{\psfig{file=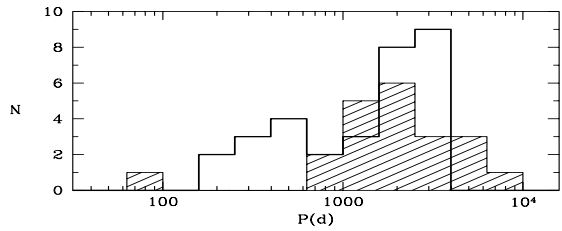,height=6cm,width=8cm}}
\end{center}
\caption[]{\label{Fig:PBa} 
The distribution of orbital periods for 21 mild barium stars 
(shaded histogram) and 31 strong barium stars (thick line) (from
McClure \& Woodsworth 1990 and Jorissen et al. 1997).
The distribution is complete up to 
about 4000~d 
}
\end{figure}

Is binarity a {\it sufficient} 
condition to produce a barium star? Probably not, 
since binary systems consisting of a {\it
normal} red giant and a WD companion with Ba-like orbital parameters
do exist (Jorissen \& Boffin 1992). 
DR Dra (= HD 160538) is probably the best example,  
with $P = 904$~d, $e = 0.07$ (compare with Fig.~\ref{Fig:elogP}) 
and a hot WD companion detected by 
Fekel et al. (1993). Berdyugina (1994) finds a metallicity 
close to solar and normal Zr and La abundances in the giant. Za\v{c}s 
et al. (1996) basically confirm that result. 

Metallicity may be the other key parameter, besides binarity, 
controlling the formation of barium stars. The s-process efficiency,
expressed in 
terms of the neutron irradiation, seems to be larger in
low-metallicity stars 
(Kov\'acs 1985; Busso et al. 1995).   
Clayton (1988) provides a theoretical foundation
for that empirical finding, provided that $^{13}\rm C(\alpha,n)^{16}O$ is 
the neutron source for the s-process. Barium stars would therefore be
easier to produce in a low-metallicity population.

\section{Inferring the mass transfer mode from the orbital 
elements: Wind accretion and/or RLOF?}

Synthetic binary evolution models (Han 
et al. 1995; de Kool \& Green 1995) 
suggest that the bimodal period distribution exhibited by 
strong barium stars (Fig.~\ref{Fig:PBa}) reflects the operation
of two distinct mass-transfer modes, RLOF in the short-period mode
(peaking around 500~d) and wind accretion 
in the long-period mode (around 3000~d). 

\begin{figure}[t]
\begin{minipage}[t]{6cm} 
\unitlength1cm
\begin{picture}(6,6)
\centerline{\psfig{file=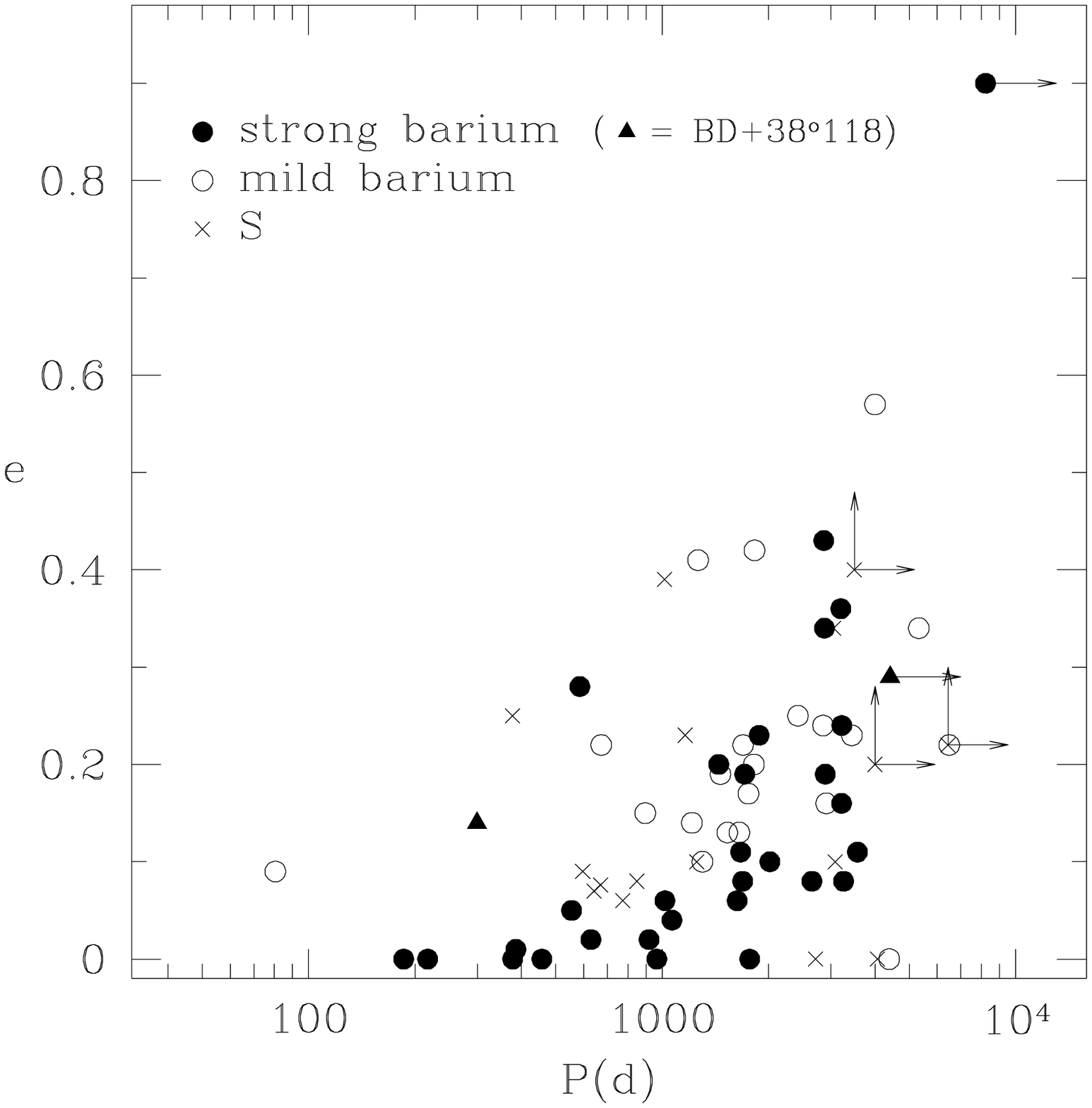,height=6cm,width=6cm}}
\end{picture}
\caption[]{\label{Fig:elogP}
The ($e, \log P$) diagram for barium and S stars (from Jorissen
et al. 1997). 
BD+38$^\circ$118 is a triple hierarchical system, with  
the close inner binary and 
the orbit of the third star around the center of mass of the inner 
binary represented by filled triangles
}
\end{minipage}
\hfill
\begin{minipage}[t]{6cm}
\begin{picture}(6,6)
\centerline{\psfig{file=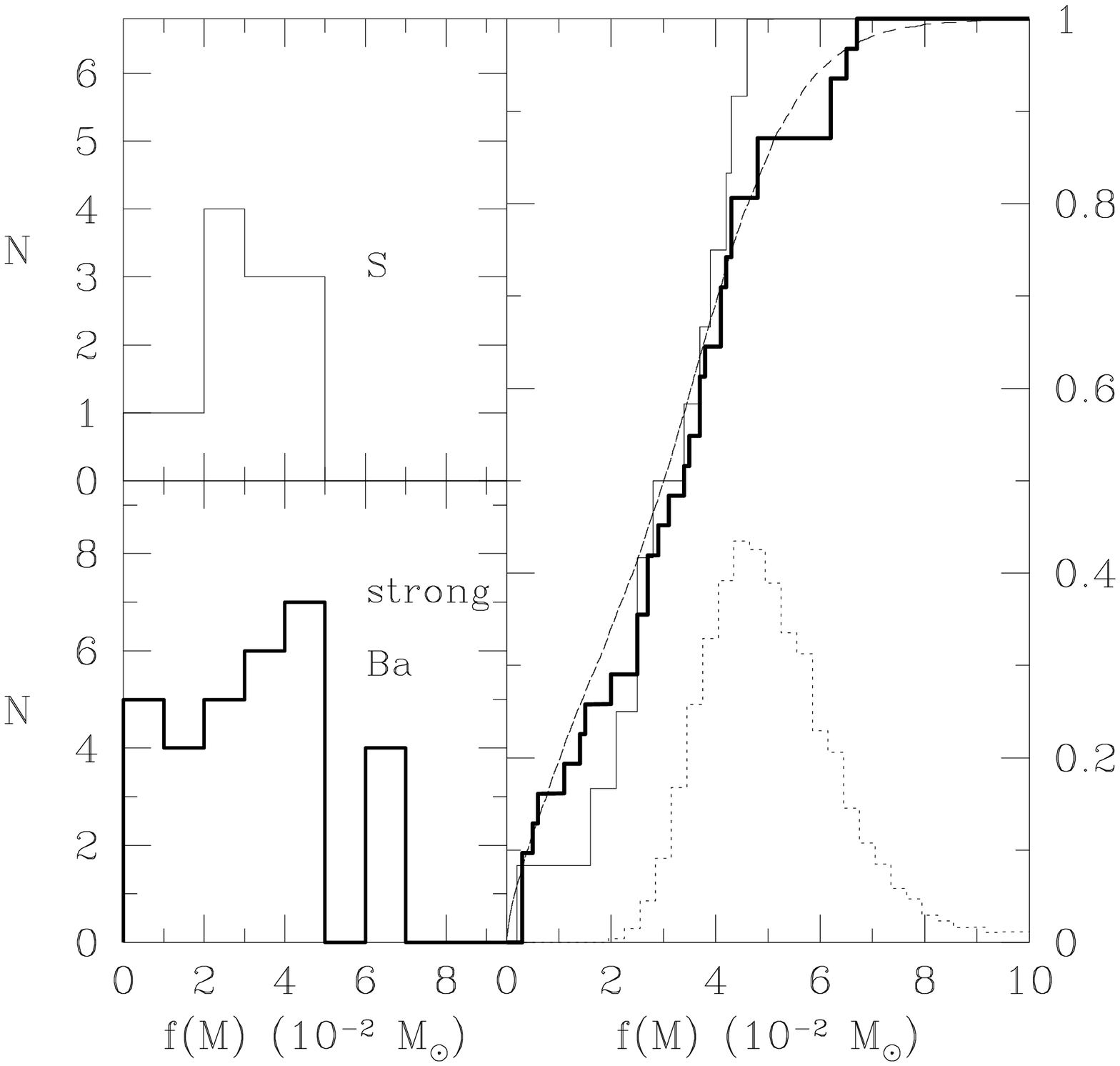,height=6cm,width=6cm}}
\end{picture}
\caption[]{\label{Fig:fMBaS}
The mass-function distributions of strong barium stars (thick line) and S 
stars (thin line) from Jorissen et al. (1997), excluding the peculiar
S systems HDE 332077 and HD 191589 (see text). The dashed line in the
right panel is the fit obtained with the $Q$ distribution shown
(dotted line)
}
\end{minipage}
\end{figure}

This general picture actually faces 
three major difficulties: first, the 
threshold period (about 1000 d) between the RLOF and 
wind-accretion modes is much too short to accomodate the large radii 
reached by AGB stars. Second, 
the period -- eccentricity diagram (Fig.~\ref{Fig:elogP}) reveals that not
all orbits in the short-period (i.e. post-RLOF) mode are circular, although 
tidal effects are expected to circularize the orbit in the phase of
large radius just preceding RLOF (e.g. Zahn 1977). A similar problem exists
for the orbits of dwarf barium stars (see North et al., this conference). 
Third, RLOF from AGB stars with a deep 
convective envelope is dynamically unstable 
(`unstable case C RLOF'; e.g. Tout \& Hall 1991),  with the ensuing 
common envelope stage generally accompanied by dramatic 
orbital shrinkage leading to the formation of a cataclysmic binary 
with a period much shorter than that of barium stars (e.g. Meyer \& 
Meyer-Hofmeister 1979). To solve these problems, Han et al. (1995), Livio
(1996) and Jorissen et 
al. (1997) propose avenues to explore. 
One of these involves Peter Eggleton's CRAP (Companion-Reinforced Attrition 
Process; Eggleton 1986) speculating that larger mass-loss 
rates for AGB stars in binary systems
may reverse the mass ratio of the system prior to RLOF, 
thus stabilizing 
the mass transfer process (Tout \& Eggleton 1988; Han et al. 1995). 

\section{Do barium stars form as dwarfs or giants?}

In Fig.~1, it  is assumed that the mass transfer responsible for the 
barium syndrome occurred when the barium star was 
still on the main sequence. Because the stellar lifetime is longer on 
the main sequence than in the giant phase, that possibility indeed 
appears more probable than the formation  of the barium star 
directly as a giant star. However, as pointed out by Iben \& Tutukov (1985),  
the mismatch between the  
thermal time scale of the dwarf's envelope and that of the mass-losing
AGB star may prevent the formation of dwarf barium stars.
A main-sequence star would indeed be driven 
out of thermal equilibrium in case of rapid mass accretion from its giant
companion, and would 
swell to giant dimensions (e.g. Kippenhahn \& Meyer-Hofmeister 1977), 
leading to a common envelope stage with 
possibly dramatic consequences on the fate of the binary system 
(see e.g. Meyer \& Meyer-Hofmeister 1979 and Sect.~2). 
Dwarf barium stars long remained elusive, until Luck \& Bond (1982, 
1991) and North et al. (1994) recognized that some of the CH 
subgiants previously identified by Bond (1974), as well as some of 
the F dwarfs previously classified by Bidelman (1985) as having 
'strong Sr $\lambda$ 4077',  
had the proper abundance anomalies, gravities and galactic 
frequencies to be identified with the long-sought Ba dwarfs. 
A large 
fraction of binaries (about 90\%) has been found among the stars with strong
anomalies, as expected (McClure 1985; 
North \& Duquennoy 1992; North et al., this conference). 
The very existence of binary dwarf Ba stars, in spite of Iben \& 
Tutukov's argument, is another indication that, if RLOF does indeed 
occur in these systems, it does not have 
the catastrophic consequences generally associated with unstable 
case C RLOF. The question of     
whether these dwarf barium stars will eventually evolve into giant 
barium stars is addressed by North et al. elsewhere in these 
Proceedings. 

The formation of a barium star directly as a giant, though probably 
less frequent, is by no means excluded. The barium star HD 
165141 may be such a case. HD 165141 is unique in sharing 
properties of barium and RS CVn systems (Fekel et al. 1993; 
Jorissen et al. 1996). Its rapid rotation ($V \sin i = 14$ \kms) and 
X-ray flux (probably from a hot corona)  are typical of RS CVn 
systems. However,  the spin-up of that star  (and the concomittant 
RS CVn properties) cannot be attributed to tidal effects 
synchronizing the stellar rotation with the orbit, as is the case for 
RS CVn systems, since the orbital period (about 5200~d) is much too 
long. That puzzle may be solved if the wind accretion episode 
responsible for the barium syndrome spun the star up, as suggested 
by detailed hydrodynamical simulations (Theuns \& Jorissen 1993; 
Theuns et al. 1996).  Since magnetic braking is generally faster than 
the stellar lifetime on the giant branch, wind accretion and 
concomittant spin-up must have occurred when HD 165141 was 
already a giant star. 
Strong support to that hypothesis comes from the fact that HD 
165141 has a hot WD companion (Fekel et al. 1993) whose cooling 
time scale is shorter than the lifetime of HD 165141 on the red 
giant branch. Finally, note that Jeffries \& Stevens (1996) have reported more 
cases of WIRRing (Wind-Induced Rapidly Rotating) stars among 
binary stars involving a hot WD.

\section{Do barium stars evolve into Tc-poor S stars?}

Figure~\ref{Fig:elogP}  shows that strong barium stars and Tc-poor S stars 
occupy the same region of the ($e, \log P$) diagram. The 
distributions  of the mass function $f(M)$ presented in
Fig.~\ref{Fig:fMBaS} [where $f(M) = M_2^3 
\sin^3 i /(M_1 + M_2)^2 \equiv Q \sin^3 i,\; M_1$ and $M_2$ being the 
masses of the giant and of the WD, respectively] for  the two 
families 
are compatible with the hypothesis that they are extracted from the
same parent population.
Following the  usual analysis (Webbink 1986; McClure \& 
Woodsworth 1990) of the mass function distribution in terms of a 
peaked distribution of mass ratios $Q$ convolved with randomly 
inclined orbits, an average ratio $Q = 0.045$ \Msun\ is found for 
the two classes, translating into a giant mass of 1.6 \Msun\ 
when adopting $M_2 = 0.6$ \Msun\ for the WD companion. These two results 
thus provide strong support to the hypothesis that barium and Tc-poor S stars 
represent successive stages in the evolutionary path sketched in Fig.~1. 

Note, however, that the above comparison of the mass functions does
not include two Tc-poor S stars (HD 191589 and HDE 332077)
with main sequence companions detected with the {\it International Ultraviolet
Explorer} satellite (Ake \& Johnson 1992; 
Ake et al. 1992). The evolutionary status of these stars is currently unknown.

\section{The relative frequency of intrinsic/extrinsic S stars}

The evaluation of the relative frequency of intrinsic/extrinsic S stars
faces two difficulties: (i) one needs an efficient criterion for 
distinguishing extrinsic from intrinsic S stars, and 
(ii) the frequency evaluation must be corrected from the selection bias,
since extrinsic and intrinsic 
S stars follow different galactic distributions (Jorissen et al. 
1993).
As far as (i) is concerned, the defining criterion  of
intrinsic/extrinsic S stars based on the presence/absence of
Tc, respectively, may be difficult to apply to 
a complete sample of S stars like Henize's (see below), since it involves many 
faint stars for which high-resolution spectroscopy is difficult to 
secure. Binarity may be an alternative, since the binary paradigm for
S stars states that all Tc-poor S stars 
should be binaries (Brown et al. 1990; Johnson 1992). 
However, some binaries must be expected  
among Tc-rich S stars as well, like in any class of stars. 
Binary intrinsic S stars with main sequence companions (case 3 in
Fig.~1) include the close visual binary $\pi^1$ Gru (Feast 1953) and  
stars with composite spectrum like T Sgr, W Aql, 
WY Cas (Herbig 1966; Culver \& Ianna 1975), and
possibly S Lyr (Merrill 1956). The situation is further confused by
extrinsic S stars reaching the AGB phase and eventually becoming Tc-rich
(case 8 in Fig.~1). $o^1$ Ori, a Tc-rich binary S star with a WD
companion (Ake \& Johnson 1988), may be such a case. 

The CORAVEL $Sb$ parameter, measuring the average line width
(see Jorissen \& Mayor 1988 for a more detailed definition), 
offers an interesting and efficient 
alternative to identify extrinsic/intrinsic S stars.

\begin{figure}
\begin{center}
\leavevmode
\centerline{\psfig{file=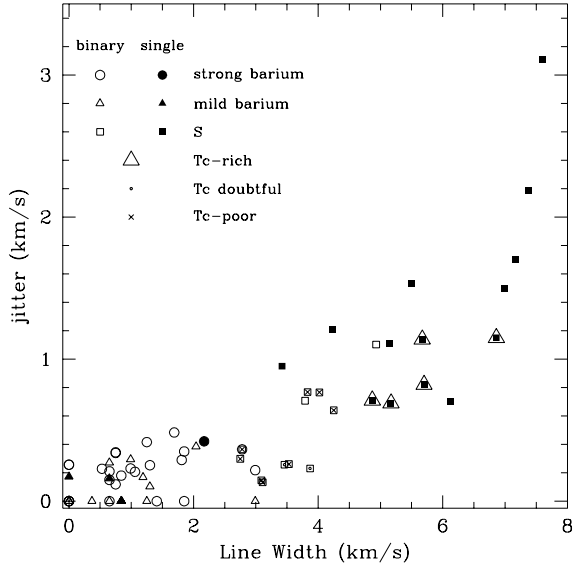,height=8cm,width=8cm}}
\end{center}
\caption[]{\label{Fig:Sbjitter}
The jitter $(\sigma^2 - \bar{\epsilon}_1^2)^{1/2}$ (where 
$\bar{\epsilon}_1$
is the average error on one measurement, and $\sigma$ is 
the standard deviation of the radial velocity for single stars, and of
the $O-C$ residuals around the computed orbit for binary stars)
vs the CORAVEL 
line-width parameter $Sb$ (see text)
}
\end{figure}

In cool red giants where macroturbulence is the main line-broadening factor, 
the $Sb$ parameter may be  expected to be a sensitive 
function of the luminosity, as is macroturbulence (e.g. Gray 1988). 
But at the same time, bright giants exhibit large
velocity jitters probably caused by envelope pulsations 
(e.g. Mayor et al. 1984).  
A correlation between $Sb$ and the radial velocity
jitter must thus be expected, as observed in Fig.~\ref{Fig:Sbjitter} for 
barium, intrinsic and extrinsic
S stars (Jorissen \& Mayor 1992; Jorissen et al. 1997). 

\begin{figure}
\begin{center}
\leavevmode
\centerline{\psfig{file=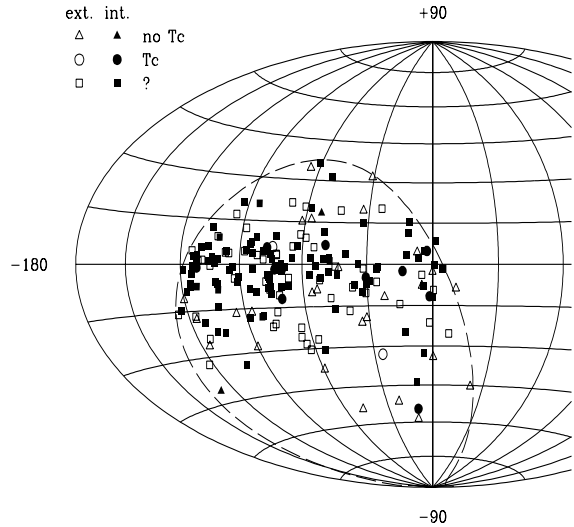,height=10cm,width=8cm}}
\end{center}
\caption[]{\label{Fig:Henize}
Galactic distribution of S stars from the Henize sample
}
\end{figure}

All Tc-poor S stars are binary 
stars, as expected, but moreover, they are restricted to $Sb <
5$~\kms. That criterion has been used to identify
extrinsic S stars among the Henize sample  (Henize 1960). 
That sample covers the sky south of declination $-25^\circ$ uniformly
to red magnitude 10.5, and 205 S stars were found from  
their ZrO $\lambda$6345 band on red-yellow
spectra with a dispersion of 450 \AA\ mm$^{-1}$ at H$\alpha$.
The galactic distribution of the Henize sample is presented in 
Fig.~\ref{Fig:Henize}.
Intrinsic S stars are clearly more concentrated towards the galactic
plane than extrinsic S stars. Correcting for the
uneven sampling of galactic latitudes, the frequency of
intrinsic S stars (based on the $Sb > 5$~\kms\ criterion) then amounts
to at least $62\pm5$\% (in a magnitude-limited sample).

\noindent {\footnotesize{\it Acknowledgements.} It is our pleasure to thank M. 
Mayor and the CORAVEL team at the Observatoire de Gen\`eve for 
making possible the long-term radial-velocity monitorings discussed here.
A.J. is Research Associate, {\it Fonds National de la Recherche 
Scientifique} (Belgium); S.V.E. is {\it Boursier F.R.I.A.} (Belgium). We 
thank the 
{\it Fonds National de la Recherche Scientifique} (Belgium), the {\it 
Communaut\'e Fran\c caise de Belgique} and the Organizing 
Committee for financial support.
}

\end{document}